\documentclass[twocolumn,showpacs,preprintnumbers]{revtex4}
\usepackage{amsmath}
\usepackage{amsfonts}
\usepackage{graphicx}
\usepackage{dcolumn}
\usepackage{bm}

\begin{document}

\title{Observing dipolar confinement-induced resonances in waveguides}
\author{ Tao Shi$^{1}$ and Su Yi$^{2}$}
\affiliation{$^{1}$Max-Planck-Institut f\"{u}r Quantenoptik, Hans-Kopfermann-Strasse. 1, 85748 Garching, Germany \\
$^{2}$State Key Laboratory of Theoretical Physics, Institute of Theoretical Physics, Chinese Academy of Sciences, P.O. Box 2735, Beijing 100190, China}
\date{\today}

\begin{abstract}
We develop a theoretical framework for the quasi-low-dimensional confinement-induced resonances (CIRs) of particles with the arbitrary three-dimensional two-body interactions, based on the Huang-Yang pseudopotential and the treatment of Feshbach resonances. Using this new approach, we analytically obtain some universal properties of dipolar CIRs in quasi-one-dimensional (1D) waveguides. We also show that the dipolar CIRs can be induced by tuning the angle between the dipole moments and the waveguide, which is experimentally observable in quasi-1D Cr and Dy atomic gases. We expect that these tilting angle induced CIRs will open up a new simpler way to control the resonant scatterings in quasi-low-dimensional systems.
\end{abstract}

\maketitle

{\em Introduction}.---Recently, ultracold gases in quasi-low-dimensional (QLD) geometries have attracted much attention~\cite{LDEx,LDEx2,LDEx3} as the reduced dimensionality leads to rich many-body physics, such as the one-dimensional (1D) Tonks-Girardeau gases~\cite{TG,TG2} and the Berezinskii-Kosterlitz-Thouless transitions in two-dimensional (2D) gases~\cite{BKT}. Remarkably, the tightly confining traps that define the QLD geometries offer an alternative means for tuning the effective two-body interactions via the confinement-induced resonances (CIRs)~\cite{CIR-1D-s}. In terms of the Feshbach resonance, the modes of the confining potential provide the open and closed channels for the QLD systems, which are coupled by the two-body interactions. A CIR occurs when the bound state energy induced by the interactions in the closed channels matches the threshold energy of the open channel~\cite{CIR-FR-s}. Over the past decade, significant progress has been made on the theoretical~\cite{CIR-FR-s,cir-general,CIR-1D-fermion,cir-suppression,Pricoupenko,cir-multichannel,CIR-inelastic,cir-coupled-l-wave} and experimental~\cite{CIR-Ex,CIR-Ex2,CIR-Ex3} studies of the CIRs with isotropic interactions.

Since the low-dimensional dipolar gases may give rise to the strongly correlated states of matter~\cite{low-dipole,low-dipole2} and the resonant dipolar Fermi gases may lead to the novel quantum phases~\cite{MCR,MCR2}, it is of great interest to study the resonant scatterings of particles with dipole-dipole interaction (DDI) in QLD geometries. The scatterings of dipolar particles in both free and confined spaces have been extensively studied~\cite{DDI-Sca,DDI-Sca2,DDI-Sca3,DDI-Sca4,DDI-Sca5,DDI-Sca6,DDI2D-Confined,DDI2D-Confined2,DDI2D-Confined3,DDI2D-Confined4,DDI-CIR-Santos,DDI-CIR-Schmelcher,DDI-CIR-Zhai}. In particular, in Refs.~\cite{DDI-CIR-Santos,DDI-CIR-Schmelcher,DDI-CIR-Zhai}, the dipolar CIRs in quasi-1D waveguides were investigated via either numerical calculations or the quantum defect theory. In general, these approaches require the truncation of the trap modes and the interaction channels, thus, do not allow for an easy generalization to study the universal properties of CIRs.

In this Letter, we develop a theoretical framework to study the confinement-induced resonant scatterings based on the Huang-Yang pseudopotential (HYPP)~\cite{HuangYang}. We first obtain the momentum representation of the HYPP, which is used to derive the explicit scattering wave function in QLD geometries by following the treatment of the Feshbach resonance. In principle, the validity of the presented theory is only limited by that of the HYPP. In addition, it is highly efficient for treating the CIRs in the presence of anisotropic interactions, as it allows one to systematically take into account the higher partial waves in the calculations.

As an application of our method, we investigate the dipolar CIRs in quasi-1D waveguide. For scatterings of bosons, we find that CIRs are completely characterized by an analytic equation which connects the resonance position, the background scattering length, and the resonance width. In particular, the resonance positions can be predicted analytically. We also show that dipolar CIRs can be induced by tuning the angle between the dipole moments and the waveguide. Remarkably, the tilting angle induced resonance is experimentally observable in quasi-1D bosonic Cr and Dy~\cite{Cr52,Dy164} gases and in the fermionic Dy gases~\cite{Dy163}.

{\em Huang-Yang pseudopotential}.---The HYPP of a general two-body interaction takes the form~\cite{Andrei}
\begin{align}
\langle{\mathbf r}|\hat{\cal V}|\Psi_{k_{0}}\rangle&=\sum_{lml^{\prime }m^{\prime }}
\frac{(2l+1)!!}{(2l^{\prime })!!} \frac{a_{lm}^{l^{\prime }m^{\prime }}(k_{0})}{2\mu k_{0}^{l+l^{\prime }}}\frac{\delta(r)}{r^{l+2}}Y_{lm}(\hat{r}) \nonumber\\
&\times\left[\partial _{r}^{2l^{\prime }+1}\left(r^{l^{\prime }+1}\int d\hat r\, Y_{l^{\prime }m^{\prime }}^{\ast }(\hat{r})\Psi _{k_{0}}(\mathbf{r})\right)\right]_{r\rightarrow 0},\nonumber
\end{align}
where $\mu$ is the reduced mass of two colliding particles, $\Psi _{k_{0}}(\mathbf{r})=\langle{\mathbf r}|\Psi_{k_{0}}\rangle$ is the scattering wave function for the particles with the incident energy $k_{0}^{2}/(2\mu)$, $a_{lm}^{l^{\prime}m^{\prime}}(k_{0})$ is the scattering length corresponding to the incoming partial wave $(lm)$ and the outgoing partial wave $(l^{\prime}m^{\prime})$, and, throughout this work, we set $\hbar=1$.  

For our purpose, it is more convenient to use the momentum representation of the HYPP. To this end, we take the Fourier transform of $\langle{\mathbf r}|\hat{\cal V}|\Psi_{k_{0}}\rangle$, which gives~\cite{suppmat}
\begin{align}
\langle {\mathbf k}|\hat{\cal V}|\Psi_{k_{0}}\rangle=\mathrm{Reg}\int d^{3}p\,
V(\mathbf{k,p})\Psi_{k_{0}}(\mathbf{p}),\nonumber
\end{align}
where $\Psi _{k_{0}}(\mathbf{k})=\langle{\mathbf k}|\Psi_{k_{0}}\rangle$, the regularization operation ${\rm Reg}(\cdot)$ picks up the regular part of the integral~\cite{Pricoupenko}, and
\begin{align}
V(\mathbf{k,p})\equiv\left\langle \mathbf{k}\right\vert\hat{\cal V}\left\vert \mathbf{p}\right\rangle
=\sum_{lml^{\prime }m^{\prime}}W_{lm}(\mathbf{k})A_{lm,l^{\prime }m^{\prime }}W_{l^{\prime }m^{\prime
}}^{\ast }(\mathbf{p}),  \nonumber
\end{align}
with $A_{lm,l^{\prime }m^{\prime }}=(\pi \mu)^{-1}i^{l^{\prime}-l}a_{lm}^{l^{\prime }m^{\prime }}(k_{0})$ representing the channel-dependent interaction strengths and $W_{lm}(\mathbf{k})\equiv\langle \mathbf{k}|W_{lm}\rangle=(k/k_{0})^{l}Y_{lm}(\hat{k})$ being the modified wave functions for the scattering channels $(lm)$. Remarkably, in terms of $|W_{lm}\rangle$, the HYPP can be formally expressed in a very compact form
\begin{eqnarray}
\hat{\mathcal V}=\sum_{lml^{\prime }m^{\prime
}}A_{lm,l^{\prime }m^{\prime }}\left\vert W_{lm}\right\rangle \left\langle
W_{l'm'}\right\vert.\label{hyppexp}
\end{eqnarray}

As an example for the HYPP, we consider the DDI between two polarized dipoles,
$V_{d}({\mathbf{r}})=d^{2}[1-3(\hat d\cdot \hat r)^{2}]/r^{3}$, where $d$ is the magnitude of the dipole moment, $\hat d$ and $\hat r$ are two unit vectors represent, respectively, the directions of the dipole moments and the position ${\mathbf r}$. Without loss of generality, we assume that $\hat d$ lies in the $xz$ plane, tilted with respect to the $z$ axis by an angle $\theta_{d}$. For this configuration, the projection of the angular momentum along the $z$ axis is not conserved unless $\theta_{d}=0$. In general, $a_{lm}^{l^{\prime }m^{\prime }}$ can only be obtained through the \textit{ab initio} calculations of the two-body scattering. However, away from shape resonances of DDI, the scattering amplitude of two colliding dipoles can be approximated by the first order Born approximation~\cite{weak dipole}, which leads to the analytic expressions for the scattering lengths $a_{lm}^{l'm'}$~\cite{suppmat}. In this case, $a_{00}^{00}$ is zero for pure DDI. Taking into account the contact interaction with the $s$-wave scattering length $a_{s}$, we have $a_{00}^{00}=a_{s}$. Now, the interaction is completely specified by three parameters: $a_{s}$, $a_{d}$ ($\equiv\mu d^{2}$, of the dimension of length), and $\theta_{d}$.

{\em Formulation}.---We consider a $(3-d)$-dimensional confining potential $U^{(3-d)}$, which divides the three-dimensional space into the $(3-d)$ confined (trapped) dimensions and the $d$ free (reduced) dimensions. Additionally, we assume that $U^{(3-d)}$ allows the separation of the center of mass and the relative motions. Using indices `$t$' and `$r$' to denote the quantities in the trapped and reduced dimensions, respectively, the Hamiltonian describing the relative motion takes the form
\begin{eqnarray}
\hat H=\hat H_{0}+\frac{\hat{\mathbf k}_{r}^{2}}{2\mu}+\hat{\cal V},\label{totH}
\end{eqnarray}
where $\hat H_{0}=\hat{\mathbf k}_{t}^{2}/(2\mu)+U^{(3-d)}({\mathbf r}_{t})$. Assuming that $|\phi_{s}\rangle$ is the eigenfunction of the confined dimensions with eigenenergy $e_{s}$, i.e.,
\begin{equation}
\hat H_{0}\left\vert \phi_{s}\right\rangle=e_{s}\left\vert\phi_{s}\right\rangle,
\end{equation}
where $s=s_{0},s_{1},...$ labels different eigenstates, and the eigenenergies are ordered as $e_{s_{0}}<e_{s_{1}}<\cdots $. If the incident energy satisfies $e_{s_{0}}<k_{0}^{2}/(2\mu)<e_{s_{1}}$, in terms of the Feshbach resonance, the scattering channel $s=s_{0}$ can be regarded as the open channel; while all other channels are the closed ones~\cite{CIR-FR-s}. We therefore define the projection operators $P=\left\vert \phi _{s_{0}}\right\rangle \left\langle \phi_{s_{0}}\right\vert$ and $Q=\sum_{s\neq s_{0}}\left\vert \phi_{s}\right\rangle \left\langle \phi _{s}\right\vert$ for the open and the closed channels, respectively. 

By using the apporach of the Feshbach resonance~\cite{FS}, the Schr\"odinger equation for the wave function in the open channel, $|\psi_{s_{0}}\rangle\equiv\langle\phi_{s_{0}}|\Psi_{k_{0}}\rangle$, becomes
\begin{equation}
\left[\hat{\mathbf k}_{r}^{2}/(2\mu)+\hat{\mathcal V}_{\rm eff}\right]\left\vert \psi
_{s_{0}}\right\rangle =q^{2}/(2\mu)\left\vert \psi
_{s_{0}}\right\rangle,\label{scheqo}
\end{equation}
where $q^{2}/(2\mu) =k_{0}^{2}/(2\mu)-e_{s_{0}}$ is the incident energy of the open channel and the effective interaction operator is
\begin{equation}
\hat{\mathcal V}_{\rm eff}=\left\langle \phi_{s_{0}}\right\vert (\hat{\mathcal V}_{PP}+\hat{\mathcal V}_{PQ}\hat{\cal G}_{Q}\hat{\mathcal V}_{QP})\left\vert \phi _{s_{0}}\right\rangle
\nonumber
\end{equation}
with $\hat{\cal V}_{PP}=P\hat{\cal V}P$, $\hat{\cal V}_{PQ}=P\hat{\cal V}Q=\hat{\cal V}_{QP}^{\dag}$, and $\hat{\cal G}_{Q}=\left[k_{0}^{2}/(2\mu)-Q\hat HQ\right]^{-1}$ being the propagator in the $Q$ space. It follows from the Lippmann-Schwinger equation that the formal scattering solution of  Eq.~(\ref{scheqo}) is
\begin{align}
|\psi_{s_{0}}^{(+)}\rangle=|{\mathbf q}\rangle+\hat{\cal G}_{0,P}\left(q^{2}/(2\mu)+i0^{+}\right)\hat{\cal T}_{\rm eff}\,|{\mathbf q}\rangle,
\end{align}
where $|{\mathbf q}\rangle$ is the incident wave, $\hat{\cal G}_{0,P}\left(q^{2}/(2\mu)\right)=\left[q^{2}/(2\mu)-\hat{\mathbf k}_{r}^{2}/(2\mu)\right]^{-1}$ is the free propagator in the $P$ space, and $$\hat{\cal T}_{\rm eff}=\hat{\cal V}_{\rm eff}+\hat{\cal V}_{\rm eff}\hat{\cal G}_{0,P}\hat{\cal T}_{\rm eff}$$ is the effective $T$-operator in the $P$ space.

To find the explicit form of $|\psi_{s_{0}}^{(+)}\rangle$, one needs to find the momentum representation of the effective interaction, namely, $V_{\mathrm{eff}}({\mathbf{k}}_{r},{\mathbf{p}}_{r})\equiv\langle{\mathbf k}_{r}|\hat{\cal V}_{\rm eff}|{\mathbf p}_{r}\rangle$. By introducing the free Green's operator in the $Q$ space, $\hat{\cal G}_{0,Q}=Q\left[k_{0}^{2}/(2\mu)-\hat H_{0}-\hat{\mathbf k}_{r}^{2}/(2\mu)\right]^{-1}Q$, it is shown straightforwardly that 
\begin{align}
V_{\mathrm{eff}}({\mathbf{k}}_{r},{\mathbf{p}}_{r})=\mathbf{w}_{s_{0}}({\mathbf{k}}_{r})\left(\mathbf{A}^{-1}-\mathbf{G}_{0,Q}\right)^{-1}\mathbf{w}_{s_{0}}^{\dag}({\mathbf{p}}_{r}),\label{veff}
\end{align}
where $\mathbf{w}_{s_{0}}$ is a {\em row} vector with components $w_{s_{0},lm}({\mathbf k}_{r})=\langle{\mathbf k}_{r}|\langle\phi_{s_{0}}|W_{lm}\rangle$, the matrix ${\mathbf A}$ is defined by the elements $A_{lm,l'm'}$, and ${\mathbf G}_{0,Q}$ by $\langle W_{lm}|\hat{\cal G}_{0,Q}|W_{l'm'}\rangle$. Subsequently, the effective $T$-matrix, $T_{\mathrm{eff}}({\mathbf{k}}_{r},{\mathbf{p}}_{r})\equiv\langle{\mathbf k}_{r}|\hat{\cal T}_{\rm eff}|{\mathbf p}_{r}\rangle$, is evaluated to be
\begin{align}
T_{\mathrm{eff}}({\mathbf{k}}_{r},{\mathbf{p}}_{r})=\mathbf{w}_{s_{0}}({\mathbf{k}}_{r})\left(\mathbf{A}^{-1}-\mathbf{G}_{0}\right)^{-1}\mathbf{w}_{s_{0}}^{\dag}({\mathbf{p}}_{r}),\label{teff}
\end{align}
where ${\mathbf G}_{0}$ is defined by the elements $\langle W_{lm}|\hat{\cal G}_{0}|W_{l'm'}\rangle$ with $\hat{\cal G}_{0}=\hat{\cal G}_{0,P}+\hat{\cal G}_{0,Q}$. These results eventually leads to the explicit form of the scattering wave function in the quasi-$d$-dimensional geometry
\begin{align}
\psi_{s_{0}}^{(+)}({\mathbf{k}}_{r})=\delta ^{(d)}({\mathbf{k}}_{r}-\mathbf{q})
+\frac{2\mu T_{\mathrm{eff}}({\mathbf{k}}_{r},{\mathbf{q}})}{q^{2}-k_{r}^{2}+i0^{+}},\label{scwav}
\end{align}
where $\psi_{s_{0}}^{(+)}({\mathbf{k}}_{r})=\langle{\mathbf k}_{r}|\psi_{s_{0}}^{(+)}\rangle$ and $\delta ^{(d)}({\mathbf{k}}_{r}-\mathbf{q})$ is the $\delta$-function in the $d$-dimensional space. Equations~(\ref{veff})-(\ref{scwav}) allow one to analyze the scattering properties systematically, which are the main results of this section.

Now, the scattering wave function in a QLD system reduces to the evaluations of ${\mathbf w}_{s_{0}}$, ${\mathbf G}_{0}$, and ${\mathbf G}_{0,Q}$, which are independent of the interaction. Fortunately, if the trapping potential takes the form of a harmonic oscillator, i.e., $U^{(2)}(x,y)=\mu\omega_{\perp}^{2}(x^{2}+y^{2})/2$ or $U^{(1)}(z)=\mu\omega_{z}^{2}z^{2}/2$ with $\omega_{\perp}$ and $\omega_{z}$ being the trap frequencies, ${\mathbf w}_{s_{0}}$, ${\mathbf G}_{0}$, and ${\mathbf G}_{0,Q}$ can be evaluated analytically (see the Supplemental material for the quasi-1D case). In what follows, we shall focus on the dipolar CIRs in quasi-1D waveguides. To this end, we adopt the width of the harmonic oscillator, $a_{\perp}=(\mu\omega_{\perp})^{-1/2}$, as the length unit. In general, ${\mathbf A}$ is a infinite-dimensional matrix for DDI. However, for practical purpose, it should be truncated by introducing a cutoff, $l_{\rm cut}$, on the angular momentum $l$. Numerically, we find that $l_{\rm cut}$ has to be larger than $28$ ($31$) for scattering in the even (odd) $l$ channels to ensure the convergence of the result. In addition, $l_{\rm cut}$ increases with growing DDI strength.

{\em Quasi-1D scatterings of dipolar bosons}.---Here, we consider the scatterings of two identical bosons in a quasi-1D waveguide, for which $l$ has to be even. In the coordinate space, the asymptotical scattering wave function becomes $\psi_{s_{0}}^{(+)}(z)=e^{iqz}-2\pi i\mu q^{-1}T_{\rm eff}(q,q)e^{iq\left\vert z\right\vert}$, from which one obtains the scattering amplitude
\begin{align}
f(q)=-\frac{2\pi i\mu }{q}T_{\rm eff}(q,q)=-\frac{2\pi i\mu V_{\rm eff}(q,q)}{q+2\pi i\mu V_{\rm eff}(q,q)}.
\end{align}
In the small $q$ limit, the scattering amplitude can be rewritten as $f(q)\rightarrow -(1+iqa_{1D})^{-1}$, where $$a_{1D}=-\frac{1}{2\pi\mu V_{\mathrm{eff}}(0,0)}$$ is the effective 1D scattering length. A CIR occurs if $a_{1D}=0$, or equivalently, $V_{\mathrm{eff}}(0,0)\rightarrow \infty $. In fact, $a_{1D}=0$ implies that the interaction between particles becomes so strong that they are impenetrable since $f(q)\rightarrow -1$. 

Using the fact that $\det {\mathbf G}_{0,Q}\neq 0$, we rewrite the inverse of the effective 1D scattering length as
\begin{equation}
a_{1D}^{-1}=2\pi\mu{\mathbf w}_{s_{0}}(0){\mathbf G}_{0,Q}^{-1}{\mathbf B}^{-1}\mathbf{A}{\mathbf w}_{s_{0}}^{\dagger }(0),
\end{equation}
where ${\mathbf B}(a_{s},a_{d})={\mathbf A}-{\mathbf G}_{0,Q}^{-1}$. Apparently, a resonance takes place when $\det {\mathbf B}=0$. To find the position of the CIRs on the $(a_{s},a_{d})$ parameter plane, we partition ${\mathbf B}$ into a block form,
\begin{equation}
{\mathbf B}=\left(
\begin{array}{cc}
b(a_{s}) & {\mathbf b}^{T}(a_{d}) \\
{\mathbf b}(a_{d}) & \widetilde {\mathbf B}(a_{d})
\end{array}
\right),\label{matrixb}
\end{equation}
where $b(a_{s})=a_{s}/(\pi\mu)-({\mathbf G}_{0,Q}^{-1})_{00,00}$ is the only element that depends on $a_{s}$, ${\mathbf b}$ is a column vector, $\widetilde{\mathbf B}$ is a square matrix, and the superscript `$T$' denotes the transpose operation. Equation (\ref{matrixb}) immediately leads to $\det {\mathbf B}=(\pi\mu)^{-1}\left(a_{s}-a_{s}^{\rm (res)}\right)\det \widetilde{\mathbf B}$, where $$a_{s}^{(\mathrm{res})}=\pi\mu\left[({\mathbf G}_{0,Q}^{-1})_{00,00}+{\mathbf b}^{T}\widetilde{\mathbf B}^{-1}{\mathbf b}\right]$$ is a polynomial of $a_{d}$, representing the position of the resonance on the $a_{s}$ axis. Apparently, for a given $a_{d}$, one can only observe a single resonance by varying $a_{s}$; while there may exist multiple resonances as one sweeps $a_{d}$ for a fixed $a_{s}$. This is exactly what has been observed in the numerical calculations~\cite{DDI-CIR-Santos}. 

To proceed further, we make use of the formal expression ${\mathbf B}^{-1}={\mathbf C}/\det {\mathbf B}$, where
\begin{equation}
{\mathbf C}=\left(
\begin{array}{cc}
c(a_{d}) & {\mathbf c}^{T}(a_{d}) \\
{\mathbf c}(a_{d}) & \widetilde{\mathbf C}(a_{s},a_{d})
\end{array}
\right)
\end{equation}
is the transpose of the cofactor matrix of ${\mathbf B}$, partitioned into the same block as Eq.~(\ref{matrixb}). It can be easily seen that $\widetilde{\mathbf C}$ linearly depends on $a_{s}$. Now, let us re-express the inverse of the effective 1D scattering length as $a_{1D}^{-1}=Z(a_{s})/\left(a_{s}-a_{s}^{(\mathrm{res})}\right)$, where $Z(a_{s})=2\left( \pi \mu \right) ^{2}(\det \widetilde{\mathbf B})^{-1}{\mathbf w}_{s_{0}}(0){\mathbf G}_{0,Q}^{-1}{\mathbf C}{\mathbf A}{\mathbf w}_{s_{0}}^{\dagger }(0)$ is a dimensionless quantity. Examination of the matrix product ${\mathbf C}{\mathbf A}$ reveals that the $Z(a_{s})$ is a linear function of $a_{s}$. Therefore, one may formally write it as $Z(a_{s})=a_{\rm bg}^{-1}\left(a_{s}-a_{s}^{(\mathrm{res})}\right)+\Delta$, which eventually leads to
\begin{equation}
\frac{1}{a_{1D}}=\frac{1}{a_{\rm bg}}+\frac{\Delta}{a_{s}-a_{s}^{(\mathrm{res})}}.\label{eqcir1d}
\end{equation}
It can be easily identified that $a_{\rm bg}$ represents the background scattering length and the dimensionless quantity $\Delta$ characterizes the width of the resonance. Not surprisingly, Eq.~(\ref{eqcir1d}) takes the similar form as that describing the Feshbach resonances~\cite{FS}. We note that, unlike $a_{s}^{({\rm res})}$, the analytic expressions for $a_{\rm bg}$ and $\Delta$ are generally unavailable. However, they can be evaluated numerically. 

\begin{figure}[t]
\includegraphics[width=0.6\columnwidth]{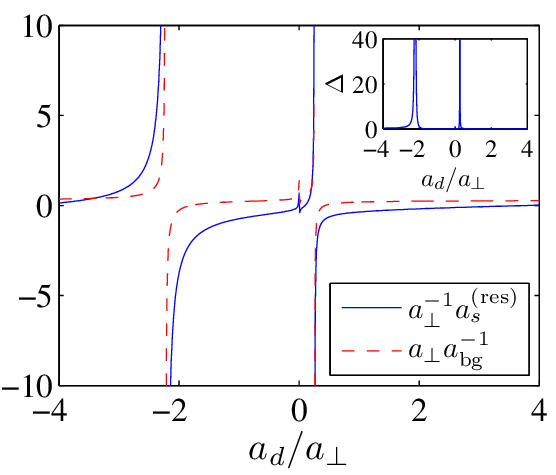}
\caption{(color online). $a_{d}$ dependence of $a_{s}^{\rm (res)}$ and $a_{\rm bg}^{-1}$. The inset shows $\Delta$ versus  $a_{d}$. The tilting angle used here is $\theta_{d}=0$.}\label{cir1d}
\end{figure}

Figure~\ref{cir1d} shows $a_{s}^{({\rm res})}$, $a_{\rm bg}^{-1}$, and $\Delta$ as functions of $a_{d}$ for $\theta_{d}=0$. It can be seen that the resonance position $a_{s}^{({\rm res})}$ moves toward the positive $a_{s}$ direction with growing $a_{d}$ and eventually diverges as $a_{d}$ approaches a critical value $a_{d}^{*}$. Once $a_{d}^{*}$ is crossed, $a_{s}^{({\rm res})}$ jumps from $+\infty$ to $-\infty$, then starts to scan the $a_{s}$ axis again until another $a_{d}^{*}$ is encountered. Similar behavior is found for $a_{\rm bg}^{-1}$. As to the resonance width, we find that $\Delta$ is always positive, in consistence with its definition as a width. In addition, $\Delta$ also diverges at $a_{d}^{*}$. In fact, $a_{s}^{({\rm res})}$, $a_{\rm bg}^{-1}$, and $\Delta$ have a same denominator, $\det\widetilde{\mathbf B}$. Hence, $a_{d}^{*}$ can be identified as the real roots of the equation $\det \widetilde{\mathbf B}=0$, where all three quantities diverge simultaneously. For $\theta_{d}\neq0$, the characteristics of the dipolar CIRs are similar to those presented in Fig.~\ref{cir1d}.

\begin{figure}[t]
\includegraphics[width=0.6\columnwidth]{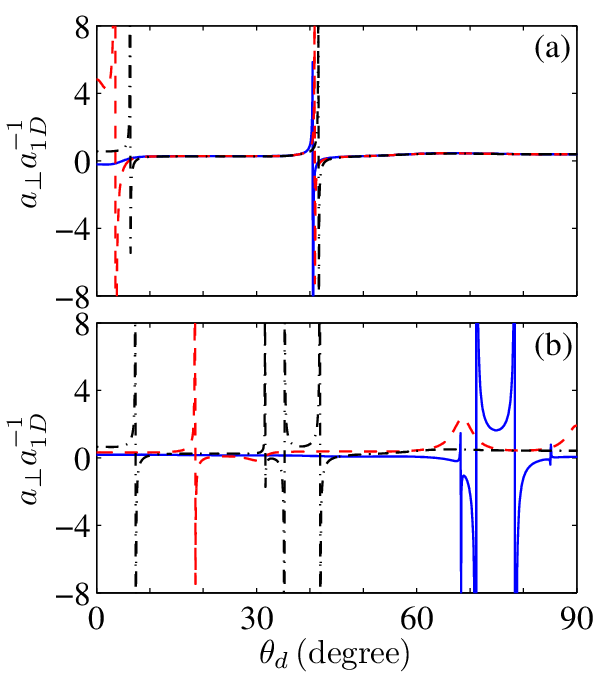}
\caption{(color online). (a) $a_{1D}^{-1}$ versus $\theta_{d}$ for $a_{s}=0a_{B}$ (solid line), $100a_{B}$ (dashed line), and $800a_{B}$ (dash-dotted line). The dipole moment is fixed at $\mu_{d}=10\mu_{B}$. (b) $a_{1D}^{-1}$ versus $\theta_{d}$ for $\mu_{d}=3.15\mu_{B}$ (solid line), $6\mu_{B}$ (dashed line), and $9\mu_{B}$ (dash-dotted line). The $s$-wave scattering length is fixed at $a_{s}=100a_{B}$.}\label{evenl}
\end{figure}

Now, we turn to investigate the $\theta_{d}$ dependence of the dipolar CIRs. Specifically, we consider the magnetic DDI with strength $d^{2}=\mu_0 \mu_{d}^{2}/(4\pi)$, where $\mu_0$ is the vacuum permeability and $\mu_{d}$ is of a few Bohr magneton ($\mu_{B}$). In addition, the mass of the atom is chosen as that of the Dy atom~\cite{Dy164} and the trap frequency is $\omega_{\perp}=(2\pi)\,14.5$kHz~\cite{CIR-Ex}. In Fig.~\ref{evenl}(a), we plot $a_{1D}^{-1}(\theta_{d})$ for various $a_{s}$ with fixed $\mu_{d}=10\mu_{B}$. As can be seen, CIRs occur for a wide range of $a_{s}$ and the positions of the resonances on the $\theta_{d}$ axis are insensitive to the variation of the $s$-wave scattering length. In Fig.~\ref{evenl}(b), we further investigate  $a_{1D}^{-1}(\theta_{d})$ with $a_{s}$ being fixed at $100a_{B}$ ($a_{B}$ is the Bohr radius). Remarkably, it is found that there exists a critical dipole moment $\mu_{d}^{*}\simeq 3.14\mu_{B}$, such that resonance always presents for $\mu_{d}\geq \mu_{d}^{*}$. It should be note that, at the critical dipole moment, the dipolar interaction strength $a_{d}$ ($\simeq 1.12\times 10^{-2}a_{\perp}$) is much smaller than the transverse confining $a_{\perp}$. Also, our calculations show that the dipolar CIRs are also observable in the bosonic Cr gases~\cite{Cr52} by tuning $\theta_{d}$.

\begin{figure}[t]
\includegraphics[width=0.6\columnwidth]{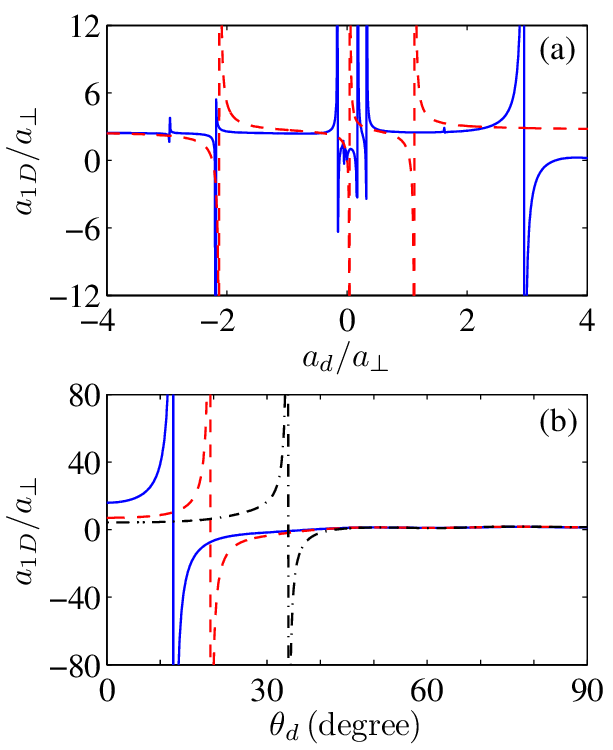}
\caption{(color online). (a) $a_{1D}$ versus $a_{d}$ for $\theta_{d}=0$ (solid line) and $\pi/2$ (dashed line). (b) $a_{1D}$ versus $\theta_{d}$ for $\mu_{d}=7\mu_{B}$ (solid line), $8\mu_{B}$ (dashed line), and $10\mu_{B}$ (dash-dotted line).}\label{oddl}
\end{figure}

{\em Quasi-1D scatterings of dipolar fermions}.---The scatterings of two identical fermions take place in the odd $l$ channels, for which the asymptotical scattering wave function is $\psi_{s_{0}}^{(+)}(z)=e^{iqz}+{\rm sign}(z)f(q) e^{iq\left\vert z\right\vert}$. For small $q$, the scattering amplitude becomes $f(q)\rightarrow -i q/(a_{1D}^{-1}+r_{1D} q^2+iq)$, where $r_{1D}$ is the effective force range and
\begin{equation}
a_{1D}=\lim_{q\rightarrow0}\frac{2\pi \mu}{q^2} V_{\mathrm{eff}}(q,q)
\end{equation}
is the effective 1D scattering length. A CIR occurs if $a_{1D}$ diverges, which also implies a total reflection.

In Fig.~\ref{oddl}(a), we show $a_{1D}$ as a function of $a_{d}$ for various $\theta_{d}$'s, which displays multiple resonances along the $a_{d}$ axis. Furthermore, Fig.~\ref{oddl}(b) plots the $\theta_{d}$ dependence of $a_{1D}$ for different $\mu_{d}$'s. Here, we have again taken $\mu$ being the mass of Dy atom~\cite{Dy163} and $\omega_{\perp}=(2\pi)\,14.5$kHz. As can be seen, for a given $\mu_{d}$, there is only one dipolar CIR within the range $\theta_{d}\in[0,\pi/2]$. In addition, the resonance position moves toward the small $\theta_{d}$ end as $\mu_{d}$ is increased. Similar to the scatterings in the even $l$ channels, CIR appears only when $\mu_{d}$ exceeds the critical value, $6.48\mu_{B}$, suggesting that the dipolar CIRs of spinless fermions is observable in the fermionic Dy gases. Finally, it is worthwhile to mention that the widths of the dipolar CIRs for fermions on the $\theta_{d}$ axis are broader than those for bosons, which is more favorable for the experimental detection.

{\em Conclusions}.---We have presented the analytic scattering wave functions in the QLD geometries for general two-body interactions describable by the HYPP. This approach allows one to systematically include all relevant partial waves in the calculation until the result is convergent. We have also studied the dipolar CIRs in quasi-1D waveguide. For bosons, we find the analytical expression for the positions of the dipolar CIRs on the axis of the $s$-wave scattering length. Strikingly, we show that, by tuning the tilting angle of the dipole moment, the dipolar CIRs is experimentally observable in quasi-1D Cr and Dy gases. Of particular importance, the introduction of the tilting angle allows the unambiguous identification of the role played by the DDI in the CIRs. 

Tao Shi thanks Carlos Navarrete Benlloch for the useful discussion. This work was supported by the EU under the IP project AQUTE, the  National 973 program under the Grants No. 2012CB922104, and NSFC (Grants No. 11025421 and No. 11121403).

\newpage
\begin{widetext}
 
\section*{Supplemental material for ``Observing dipolar confinement-induced resonances in waveguides''}

\subsection{Momentum representation of the Huang-Yang pseudopotential}
To derive the momentum representation of the HYPP, we consider the Fourier transform of $\hat{\cal V}\Psi_{k_{0}}({\mathbf r})$, i.e.,
\begin{align}
\int\frac{d^{3}r}{(2\pi)^{3/2}}e^{-i{\mathbf k}\cdot{\mathbf r}}\hat{\cal V}\Psi_{k_{0}}({\mathbf r})=&\langle {\mathbf k}|\hat{\cal V}|\Psi_{k_{0}}\rangle\nonumber\\
=&\sum_{lml^{\prime }m^{\prime }}\frac{(2l+1)!!}{(2l^{\prime })!!}\frac{a_{lm}^{l^{\prime }m^{\prime
}}(k_{0})}{2\mu k_{0}^{l+l^{\prime }}}\!\int\!
\frac{d^{3}r}{(2\pi )^{3/2}}e^{-i\mathbf{k\cdot r}}\frac{\delta(r)}{r^{l+2}}Y_{lm}(\hat{r})\nonumber\\
&\times\left[ \partial _{r}^{2l^{\prime }+1}\left( r^{l^{\prime }+1}\!\int\! d
\hat{r}Y_{l^{\prime }m^{\prime }}^{\ast }(\hat{r})\!\int\!\frac{d^{3}k'}{(2\pi )^{3/2}}e^{i{\mathbf k}'\cdot {\mathbf r}}\Psi _{k_{0}}(\mathbf{k}')\right) \right] _{r\rightarrow 0}.\label{HYPPp1}
\end{align}
To simplify Eq. (\ref{HYPPp1}), we make use of the expansions 
\begin{eqnarray}
e^{i{\mathbf k}\cdot{\mathbf r}}=4\pi\sum_{lm}i^{l}j_{l}(kr)Y_{lm}^{*}(\hat{k})Y_{lm}(\hat r). \label{sphexp}
\end{eqnarray}
In addition, the $\delta$-function in the first integral allows us to use the asymptotical expression of $j_{l}(kr)$ at $r\rightarrow0$. As a result, Eq.~(\ref{HYPPp1}) reduces to
\begin{align}
\langle {\mathbf k}|\hat{\cal V}|\Psi_{k_{0}}\rangle=\frac{1}{\pi\mu}\sum_{lml'm'}\left(\frac{k}{k_{0}}\right)^{l}Y_{lm}(\hat{k})i^{l'-l}a_{lm}^{l'm'}(k_{0})\frac{1}{k_{0}^{l'}(2l')!!}\left[\partial_{r}^{2l'+1}\!\left(r^{l'+1}\int d^{3}k'\,j_{l'}(k'r)Y_{l'm'}^{*}(\hat k')\Psi_{k_{0}}({\mathbf k}')\right)\right]_{r\rightarrow0}.\label{HYPPp2}
\end{align}
Replacing $|\Psi_{k_{0}}\rangle$ by $|{\mathbf p}\rangle$, we find
\begin{align}
\langle {\mathbf k}|\hat{\cal V}|{\mathbf p}\rangle=\frac{1}{\pi\mu}\sum_{lml'm'}\left(\frac{k}{k_{0}}\right)^{l}Y_{lm}(\hat{k})i^{l'-l}a_{lm}^{l'm'}(k_{0})\frac{1}{k_{0}^{l'}(2l')!!}\left[\partial_{r}^{2l'+1}\!\left(r^{l'+1}j_{l'}(pr)Y_{l'm'}^{*}(\hat p)\right)\right]_{r\rightarrow0}.
\label{HYPPp3}
\end{align}
Using the asymptotical expression of $j_{l}(pr)$ at $r\rightarrow0$, a straightforward calculation immediately leads to
\begin{align}
V(\mathbf{k,p})\equiv\left\langle \mathbf{k}\right\vert\hat{\cal V}\left\vert \mathbf{p}\right\rangle
=\frac{1}{\pi \mu}\sum_{lml^{\prime }m^{\prime}}W_{lm}(\mathbf{k})i^{l^{\prime}-l}a_{lm}^{l^{\prime }m^{\prime }}(k_{0})W_{l^{\prime }m^{\prime
}}^{\ast }(\mathbf{p}).
\end{align}

\subsection{Scattering lengths of DDI}
Here, we derive the scattering lengths of the DDI away from the shape resonance by the first order Born approximation (FBA). Generally, we assume that the direction of the dipole moment is
\begin{equation}
\hat{d}=(\sin \theta _{d}\cos \varphi _{d},\sin \theta_{d}\sin \varphi _{d},\cos \theta _{d}),
\end{equation}
where $\theta_{d}$ and $\varphi_{d}$ are, respectively, the polar and azimuthal angles of the dipole moment. Using the spherical harmonic functions, the DDI, $V_{d}(\mathbf{r})=d^{2}[1-3(\hat{d}\cdot \hat{r})^{2}]/r^{3}$, can be rewritten as
\begin{equation}
V_{d}(\mathbf{r})=-\frac{d^{2}}{r^{3}}\sqrt{\frac{\pi }{5}}%
\sum_{q=-2}^{2}v_{q}Y_{2q}(\hat{r}),
\end{equation}%
where $v_{\pm 2}=\sqrt{6}\,\hat{d}_{\mp }^{2}$, $v_{\pm 1}=\mp 2\sqrt{6}\,\hat{d}_{\mp }\hat{d}_{z}$, and $v_{0}=4\left(\hat{d}_{z}^{2}-\hat{d}_{+}\hat{d}_{-}/2\right)$ with $\hat{d}_{\pm }=\sin \theta _{d}e^{\pm i\varphi _{d}}$ and $\hat{d}_{z}=\cos \theta_{d}$. With the FBA, the scattering lengths in the mixed channels $lm$ and $l^{\prime}m^{\prime }$ are given by
\begin{equation}
a_{lm}^{l^{\prime }m^{\prime }}=2\mu \lim_{k\rightarrow p}\int \frac{d\hat{k}d\hat{p}}{(4\pi )^{2}}i^{l-l^{\prime }}Y_{lm}^{\ast }(\hat{k})Y_{l^{\prime }m^{\prime }}(\hat{p})V_{d}(\mathbf{k}-\mathbf{p}).
\end{equation}
where $V_{d}(\mathbf{k}-\mathbf{p})=\int d^{3}rV_{d}(\mathbf{r})e^{-i(\mathbf{k}-\mathbf{p})\cdot \mathbf{r}}$
is the Fourier transform of $V_{d}(\mathbf{r})$. The expansion Eq. (\ref{sphexp}) gives
\begin{eqnarray}
a_{lm}^{l'm'}=-2\sqrt{\frac{\pi }{5}}a_{d}\int_{0}^{%
\infty }dr\frac{1}{r}j_{l}(kr)j_{l^{\prime }}(kr)\sum_{q=-2}^{2}v_{q}\int d\hat r\, Y_{lm}^{\ast }(\hat{r}%
)Y_{2,q}(\hat{r})Y_{l^{\prime }m^{\prime }}(\hat{r}).
\end{eqnarray}%
The two integrals in the above equation can be evaluated explicitly, namely,
\begin{equation}
\int_{0}^{\infty }dx\frac{1}{x}j_{l}(x)j_{l^{\prime }}(x)=-\frac{2\cos \left(
\frac{l-l^{\prime }}{2}\pi \right)}{[(l-l^{\prime })^{2}-1](l+l^{\prime
})(l+l^{\prime }+2)}
\end{equation}%
and%
\begin{eqnarray}
\int d\hat r\, Y_{lm}^{\ast }(\hat{r})Y_{2,q}(\hat{r})Y_{l^{\prime
}m^{\prime }}(\hat{r})=(-1)^{m}\sqrt{\frac{5(2l+1)(2l^{\prime }+1)}{4\pi }}\left(
\begin{array}{ccc}
l & 2 & l^{\prime } \\
-m & q & m^{\prime }%
\end{array}%
\right) \left(
\begin{array}{ccc}
l & 2 & l^{\prime } \\
0 & 0 & 0
\end{array}
\right),
\end{eqnarray}
which eventually lead to
\begin{eqnarray}
a_{lm}^{l'm'}=\frac{2(-1)^{m}a_{d}\cos \left(\frac{l-l^{\prime
}}{2}\pi \right)\sqrt{(2l+1)(2l^{\prime }+1)}}{[(l-l^{\prime })^{2}-1](l+l^{\prime
})(l+l^{\prime }+2)} \sum_{q=-2}^{2}v_{q}\left(
\begin{array}{ccc}
l & 2 & l^{\prime } \\
-m & q & m^{\prime }%
\end{array}%
\right) \left(
\begin{array}{ccc}
l & 2 & l^{\prime } \\
0 & 0 & 0%
\end{array}%
\right) .
\end{eqnarray}

\subsection{Derivations of ${\mathbf w}_{s}$ and $\mathbf{G}_{0,Q}$ in quasi-1D waveguide}
The eigenfunction of the 2D harmonic oscillator, $U^{(2)}(x,y)=\mu\omega_{\perp}(x^{2}+y^{2})/2$, is
\begin{eqnarray}
\phi _{nm}(k_{\rho },\varphi_{k})=(-1)^{n-\left\vert m\right\vert }
\sqrt{\frac{[(n+\left\vert m\right\vert)/2]!}{[(n-\left\vert m\right\vert)/2]!}}\frac{a_{\perp }}{\sqrt{\pi }\left\vert m\right\vert !}(k_{\rho }a_{\perp })^{\left\vert m\right\vert } {}_{1}F_{1}\left(\frac{\left\vert m\right\vert -n}{2},\left\vert m\right\vert +1,k_{\rho }^{2}a_{\perp }^{2}\right)\,e^{-\frac{1}{2}k_{\rho }^{2}a_{\perp }^{2}}e^{im\varphi _{k}},
\end{eqnarray}
where $k_{\rho}=\sqrt{k_{x}^{2}+k_{y}^{2}}$, $\varphi_{k}$ is the azimuthal angle of ${\mathbf k}$, $_{p}F_{q}(a_{1},...,a_{p};b_{1},...b_{q};z)$ is the generalized hypergeometric function, and $n=2n_{\rho}+|m|$ with $n_{\rho}=0,1,2,\ldots$ and $m=0,\pm1,\pm2,\ldots$. The corresponding eigenenergy is $e_{n}=(n+1)\omega_{\perp}$.

Let us first evaluate ${\mathbf w}_{s}$. In quasi-1D waveguide, the component of ${\mathbf w}_{s}$ is
\begin{equation}
w_{nm',lm}(k_{z})=\left\langle \phi_{nm'}\left\vert W_{lm}\right\rangle \right.
=\int d^{2}k_{\rho }\phi_{nm'}^{\ast }(k_{\rho })W_{lm}({\mathbf k})=\int d^{2}k_{\rho }\phi _{nm'}^{\ast }(k_{\rho })\left( \frac{k}{k_{0}}
\right) ^{l}Y_{lm}(\hat k),\label{wnlm}
\end{equation}
which is nonzero only for $m=m'$. It can therefore be denoted as $w_{n,lm}$. To evaluate the above integral, we employ the definition of the spherical harmonic function, i.e.,
\begin{eqnarray}
Y_{lm}(\hat{k})={\rm sign}^{m}(-m)\sqrt{\frac{2l+1}{4\pi }\frac{(l-\left\vert
m\right\vert )!}{(l+\left\vert m\right\vert)!}}\,e^{im\varphi _{k}}(1-\cos ^{2}\theta _{k})^{\left\vert m\right\vert /2}\left. \partial
_{x}^{\left\vert m\right\vert }P_{l}(x)\right\vert _{x=\cos \theta _{k}},
\end{eqnarray}
where 
\begin{equation}
P_{l}(x)=\frac{1}{2^{l}}%
\sum_{j=0}^{[l/2]}(-1)^{j}C_{l}^{j}C_{2l-2j}^{l}x^{l-2j}
\end{equation}
is the Legendre function with $C_{n}^{k}\equiv\left(\begin{array}{c}n\\k\end{array}\right)$ being the binomial coefficients and $[x]$ denoting the largest integer not greater than $x$. With this expansion, $W_{lm}({\mathbf k})$ can be rewritten as
\begin{eqnarray}
W_{lm}({\mathbf k})=\frac{{\rm sign}^{m}(-m)}{2^{l}k_{0}^{l}}\sqrt{\frac{2l+1}{4\pi }
\frac{(l-\left\vert m\right\vert )!}{(l+\left\vert m\right\vert )!}}\,
e^{im\varphi _{k}}\sum_{j^{\prime }=0}^{\left[\frac{l-\left\vert m\right\vert }{2}\right]}
\sum_{j=j^{\prime }}^{\left[\frac{l-\left\vert m\right\vert }{2}\right
]}(-1)^{j}C_{l}^{j}C_{2l-2j}^{l}C_{j}^{j^{\prime }}\frac{(l-2j)!}{%
(l-\left\vert m\right\vert -2j)!} k_{\rho }^{2j^{\prime }+\left\vert
m\right\vert }k_{z}^{l-\left\vert m\right\vert -2j^{\prime }}.\nonumber
\end{eqnarray}
Here, the summation over $j$ can be performed by employing the result
\begin{eqnarray}
\sum_{j=j^{\prime }}^{\left[\frac{l-\left\vert m\right\vert }{2}%
\right]}(-1)^{j}C_{l}^{j}C_{2l-2j}^{l}C_{j}^{j^{\prime }}\frac{(l-2j)!}{%
(l-\left\vert m\right\vert -2j)!}=\frac{(-1)^{j^{\prime }}2^{l-\left\vert m\right\vert -2j^{\prime }}\Gamma
(l+\left\vert m\right\vert +1)}{\Gamma (j^{\prime }+\left\vert m\right\vert
+1)\Gamma (j^{\prime }+1)\Gamma (l-\left\vert m\right\vert -2j^{\prime }+1)},
\end{eqnarray}
which leads to
\begin{eqnarray}
W_{lm}({\mathbf k}) =\frac{{\rm sign}^{m}(-m)}{2^{l}k_{0}^{l}}\sqrt{\frac{2l+1}{4\pi }%
\frac{(l-\left\vert m\right\vert )!}{(l+\left\vert m\right\vert )!}}\,
e^{im\varphi _{k}}\sum_{j^{\prime }=0}^{\left[\frac{l-\left\vert m\right\vert }{2}\right]}\frac{%
(-1)^{j^{\prime }}2^{l-\left\vert m\right\vert -2j^{\prime }}\Gamma
(l+\left\vert m\right\vert +1)k_{\rho }^{2j^{\prime }+\left\vert
m\right\vert }k_{z}^{l-\left\vert m\right\vert -2j^{\prime }}}{\Gamma
(j^{\prime }+\left\vert m\right\vert +1)\Gamma (j^{\prime }+1)\Gamma
(l-\left\vert m\right\vert -2j^{\prime }+1)}.
\end{eqnarray}
Furthermore, the integral in Eq. (\ref{wnlm}) can be worked out analytically by using the formula
\begin{eqnarray}
\int_{0}^{\infty }dze^{-z}z^{s-1}\text{ }_{p}F_{q}(a_{1},...,a_{p};b_{1},...,b_{q};az)=\Gamma (s)\text{ }_{p+1}F_{q}(s,a_{1},...,a_{p};b_{1},...,b_{q};a),
\end{eqnarray}
which eventually leads to
\begin{eqnarray}
w_{n,lm}(k_{z}) &=&\frac{(-1)^{n-\left\vert m\right\vert }{\rm sign}^{m}(-m)}{%
\left\vert m\right\vert !(k_{0}a_{\perp })^{l}a_{\perp }}\sqrt{(2l+1)\frac{%
(l-\left\vert m\right\vert )!}{(l+\left\vert m\right\vert )!}}\sqrt{\frac{[(n+\left\vert m\right\vert)/2]!}{[(n-\left\vert m\right\vert)/2]!}}  \notag \\
&&\times\sum_{j^{\prime }=0}^{\left[\frac{l-\left\vert m\right\vert }{2}\right]}\frac{\left(-\frac{
1}{2}\right)^{j^{\prime }}\Gamma (l+\left\vert m\right\vert +1)}{\Gamma (j^{\prime
}+1)\Gamma (l-\left\vert m\right\vert -2j^{\prime }+1)}\text{ }_{2}F_{1}\left(j^{\prime }+\left\vert m\right\vert +1,\frac{\left\vert m\right\vert -n}{2};\left\vert m\right\vert +1;2\right)(k_{z}a_{\perp
})^{l-\left\vert m\right\vert -2j^{\prime }}.  \label{w}
\end{eqnarray}

Finally, we evaluate the free Green's function in the $Q$ space. Since $\hat H_{0}$ conserves the projection of angular momentum along the $z$ axis, $\mathbf{G}_{0,Q}$ is diagonal with respect to the index $m$, i.e., $\left( \mathbf{G}_{0,Q}\right) _{lm,l^{\prime }m^{\prime }}=\left( \mathbf{G}_{0,Q}\right) _{lm,l^{\prime}m}\delta_{mm^{\prime }}$. Following its definition, we decompose $\left( \mathbf{G}_{0,Q}\right) _{lm,l^{\prime}m}$ into
\begin{eqnarray}
\left( \mathbf{G}_{0,Q}\right) _{lm,l^{\prime }m^{\prime }}=\delta_{m0}\sum_{n_{\rho }=1}^{\infty }\int_{-\infty }^{+\infty }dk_{z}\frac{w_{2n_{\rho},l0}^{\ast }(k_{z})w_{2n_{\rho},l^{\prime }0}(k_{z})}{\frac{q^{2}}{2\mu }-\frac{k_{z}^{2}}{2\mu }-2n_{\rho }\omega _{\perp }}+(1-\delta _{m0})\sum_{n_{\rho}=0}^{\infty }\int dk_{z}\frac{w_{n,lm}^{\ast }(k_{z})w_{n,l^{\prime }m}(k_{z})}{\frac{q^{2}}{2\mu }-\frac{k_{z}^{2}}{2\mu }-(2n_{\rho }+\left\vert m\right\vert )\omega _{\perp }}.\label{gdef}
\end{eqnarray}%
where $q^{2}/2\mu <\omega _{\perp }$. Using Eq. (\ref{w}), the integrals in Eq. (\ref{gdef}) (independent of the value of $m$) can be evaluated as
\begin{align}
\int_{-\infty }^{+\infty }dk_{z}\frac{w_{n,lm}^{\ast
}(k_{z})w_{n,l^{\prime }m}(k_{z})}{\frac{q^{2}}{2\mu }-\frac{k_{z}^{2}}{2\mu }%
-(2n_{\rho }+\left\vert m\right\vert )\omega _{\perp }}=&-\frac{(-4)^{\frac{l+l^{\prime }}{2}-\left\vert m\right\vert }\pi \mu }{%
\left\vert m\right\vert !^{2}(k_{0}a_{\perp })^{l+l^{\prime }}a_{\perp }}%
\sqrt{(2l+1)(2l^{\prime }+1)\frac{(l-\left\vert m\right\vert )!}{%
(l+\left\vert m\right\vert )!}\frac{(l^{\prime }-\left\vert m\right\vert )!}{%
(l^{\prime }+\left\vert m\right\vert )!}}  \notag \\
&\times\sum_{j=0}^{[\frac{l-\left\vert m\right\vert }{2}]}\sum_{j^{\prime }=0}^{[%
\frac{l^{\prime }-\left\vert m\right\vert }{2}]}\frac{(-\frac{1}{2}%
)^{3j+3j^{\prime }}\Gamma (l+\left\vert m\right\vert +1)\Gamma (l^{\prime
}+\left\vert m\right\vert +1)}{\Gamma (j+1)\Gamma (l-\left\vert m\right\vert
-2j+1)\Gamma (j^{\prime }+1)\Gamma (l^{\prime }-\left\vert m\right\vert
-2j^{\prime }+1)}  \notag \\
&\times \left(n_{\rho }+\frac{\left\vert m\right\vert }{2}-\tau
\right)^{\frac{l+l^{\prime }}{2}-\frac{1}{2}-\left\vert m\right\vert -j-j^{\prime
}}(-1)^{j+j^{\prime }}\frac{(n_{\rho }+\left\vert m\right\vert )!}{n_{\rho }!}  \notag \\
&\times\text{ }_{2}F_{1}(j+\left\vert m\right\vert +1,-n_{\rho };\left\vert
m\right\vert +1;2)\text{ }_{2}F_{1}(j^{\prime }+\left\vert m\right\vert
+1,-n_{\rho };\left\vert m\right\vert+1;2).
\end{align}%
where $\tau =q^{2}a_{\perp }^{2}/4$. To calculate the summation over $n_{\rho }$ in Eq. (\ref{gdef}), we introduce the function $D_{j,j^{\prime },\left\vert m\right\vert}^{j^{\prime \prime }}(\cdot)$ by%
\begin{eqnarray*}
&&(-1)^{j+j^{\prime }}\frac{(n_{\rho }+\left\vert m\right\vert )!}{n_{\rho }!%
}\text{ }_{2}F_{1}(j+\left\vert m\right\vert +1,-n_{\rho };\left\vert
m\right\vert +1;2)\text{ }_{2}F_{1}(j^{\prime }+\left\vert m\right\vert
+1,-n_{\rho };\left\vert m\right\vert +1;2) \\
&=&\sum_{j^{\prime \prime }=0}^{\left\vert m\right\vert +j+j^{\prime
}}D_{j,j^{\prime },\left\vert m\right\vert }^{j^{\prime \prime }}\left(\tau -%
\frac{\left\vert m\right\vert }{2}\right)\left(n_{\rho }+\frac{\left\vert m\right\vert
}{2}-\tau\right)^{j^{\prime \prime }},
\end{eqnarray*}%
which is the polynomial of $n_{\rho }$ with the highest order $n_{\rho}^{\left\vert m\right\vert +j+j^{\prime }}$. By using the Zeta function regularization
\begin{equation}
\xi _{\mathrm{reg}}(s,\tau )=\sum_{n=1}^{\infty }\frac{1}{(n+\tau )^{s}}=\xi
(s,\tau )-\frac{1}{\tau ^{s}}
\end{equation}%
and
\begin{equation}
\xi (s,\tau )=\sum_{n=0}^{\infty }\frac{1}{(n+\tau )^{s}},
\end{equation}%
we obtain the closed channel Green functions
\begin{eqnarray}
\left( \mathbf{G}_{0,Q}\right) _{l0,l^{\prime }0} &=&-\frac{\pi \mu (-4)^{%
\frac{l+l^{\prime }}{2}}}{(k_{0}a_{\perp })^{l+l^{\prime }}a_{\perp }}\sqrt{%
(2l+1)(2l^{\prime }+1)}  \notag \\
&&\times\sum_{j=0}^{[\frac{l}{2}]}\sum_{j^{\prime }=0}^{[\frac{l^{\prime }}{2}%
]}\sum_{j^{\prime \prime }=0}^{j+j^{\prime }}\frac{(-\frac{1}{2}%
)^{3j+3j^{\prime }}\Gamma (l+1)\Gamma (l^{\prime }+1)D_{j,j^{\prime
},0}^{j^{\prime \prime }}(\tau )}{\Gamma (j+1)\Gamma (l-2j+1)\Gamma
(j^{\prime }+1)\Gamma (l^{\prime }-2j^{\prime }+1)}  \notag \\
&&\times\,\xi _{\mathrm{reg}}\left(\frac{1}{2}+j+j^{\prime }-j^{\prime \prime }-\frac{%
l+l^{\prime }}{2},-\tau \right)
\end{eqnarray}%
for $m=0$ and%
\begin{eqnarray}
\left( \mathbf{G}_{0,Q}\right) _{lm,l^{\prime }m} &=&-\frac{\pi\mu(-4)^{\frac{%
l+l^{\prime }}{2}-\left\vert m\right\vert }}{\left\vert m\right\vert
!^{2}(k_{0}a_{\perp })^{l+l^{\prime }}a_{\perp }}\sqrt{(2l+1)(2l^{\prime }+1)%
\frac{(l-\left\vert m\right\vert )!}{(l+\left\vert m\right\vert )!}\frac{%
(l^{\prime }-\left\vert m\right\vert )!}{(l^{\prime }+\left\vert
m\right\vert )!}}  \notag \\
&&\times\sum_{j=0}^{[\frac{l-\left\vert m\right\vert }{2}]}\sum_{j^{\prime }=0}^{[%
\frac{l^{\prime }-\left\vert m\right\vert }{2}]}\sum_{j^{\prime \prime
}=0}^{\left\vert m\right\vert +j+j^{\prime }}\frac{(-\frac{1}{2}%
)^{3j+3j^{\prime }}\Gamma (l+\left\vert m\right\vert +1)\Gamma (l^{\prime
}+\left\vert m\right\vert +1)D_{j,j^{\prime },\left\vert m\right\vert
}^{j^{\prime \prime }}\left(\tau -\frac{\left\vert m\right\vert }{2}\right)}{\Gamma
(j+1)\Gamma (l-\left\vert m\right\vert -2j+1)\Gamma (j^{\prime }+1)\Gamma
(l^{\prime }-\left\vert m\right\vert -2j^{\prime }+1)}  \notag \\
&&\times\,\xi \left(\frac{1}{2}+\left\vert m\right\vert +j+j^{\prime }-j^{\prime \prime }-%
\frac{l+l^{\prime }}{2},\frac{\left\vert m\right\vert }{2}-\tau\right)
\end{eqnarray}%
for $m\neq 0$.

\end{widetext}
\end{document}